 \documentclass[onecolumn,12pt,draft]{IEEEtran} 

\usepackage{epsf}

\begin{document}

\title{ 
From Sigmoid Power Control Algorithm to Hopfield-like Neural Networks:
``SIR'' (``Signal''-to-``Interference''-Ratio)-
Balancing Sigmoid-Based Networks- Part I: Continuous Time
} 

\author{Zekeriya Uykan \thanks{Z. Uykan is with Helsinki University of Technology, Control 
Engineering Laboratory, FI-02015 HUT, Finland. E-mail: zekeriya.uykan@hut.fi. 
The author is a visiting scientist at Harvard University Broadband Comm Lab., Cambridge, MA, 
and this work has been performed during his stay at Harvard University.  
} 
}


\maketitle
\begin{abstract}
Continuous-time Hopfield network has been an important focus of 
research area since 1980s whose applications vary from image restoration to 
combinatorial optimization 
from control engineering to associative memory systems.  
On the other hand, in wireless communications systems literature, 
power control has been intensively studied as an essential mechanism for increasing the system performance. 
A fully distributed power control algorithm (DPCA), called Sigmoid DPCA, is presented by Uykan in 
\cite{UykanPhD01} and \cite{Uykan04}, 
which is obtained  by discretizing the continuous-time system.
In this paper, we present a Sigmoid-based ``Signal-to-Interference Ratio, (SIR)'' balancing
dynamic networks, called Sgm''SIR''NN, which includes both the Sigmoid power control algorithm (SgmDPCA) and the 
Hopfield neural networks, two different areas whose scope of interest, motivations and settings 
are completely different. It's shown that the Sgm''SIR''NN exhibits features which are generally 
attributed to Hopfield Networks. Computer simulations show the effectiveness of the 
proposed network as compared to traditional Hopfield Network.

\end{abstract}


\section{Introduction \label{Section:INTRO}}

\PARstart{I}{n} neural networks field, continuous-time Hopfield Neural Networks \cite{Hopfield85} 
has been an important focus of 
research area since early 1980s whose applications vary from combinatorial 
optimization (e.g. \cite{Matsuda98}, \cite{Smith98} among many others) 
including traveling salesman problem (e.g. \cite{Tan05}, \cite{Huajin04} among others) 
to image restoration (e.g. \cite{Paik92}), 
from various control engineering optimization problems including in robotics 
(e.g. \cite{Lendaris99} among others) to associative memory systems 
(e.g. \cite{Farrel90}, \cite{Muezzinoglu05} among others), etc. 
For a tutorial and further references about Hopfield NN, see e.g. \cite{Zurada92},  
\cite{Haykin99}, \cite{Vidyasagar93}.

On the other hand, in the cellular radio systems literature, power control has been 
intensively studied as an essential mechanism for high-capacity cellular networks. 
In this paper, we construct a bridge between two different areas, 
Hopfield-like Neural Networks and fully distributed power control algorithms,  
whose scope of interest, motivations and settings are completely different.

Transmitter power control is essential for
high-capacity cellular radio systems \cite{Zander92a}, \cite{Rappaport96}, etc. Power Control (PC)
problem has drawn much attention since Zander's works on centralized \cite{Zander92a} and
distributed \cite{Zander92b} {\it CIR balancing}. Carrier-to-Interference+noise Ratio (CIR) 
balancing was further investigated by Grandhi {\it et al.} \cite{Grandhi93},
\cite{Grandhi94a}. In \cite{Foschini93}, Foschini and Miljanic
considered a more general and realistic model, in which a positive
receiver noise and a respective target SIR were taken into
account. The Foschini and Miljanic's distributed algorithm (FMA)
was shown to converge either synchronously \cite{Foschini93} or
asynchronously \cite{Mitra93} to a fixed point of a {\it feasible}
system. Based on the FMA, Grandhi {\it et al.} \cite{Grandhi95}
suggested {\it distributed constrained power control} (DCPC) algorithm, in
which a transmission upper limit was considered. Some papers focus
on convergence speed of the PC operation e.g. \cite{Leung96}, \cite{riku99a},
\cite{Uykan00}, \cite{{Lv00}}, \cite{Uykan04}.  It would be very difficult to prepare a complete list 
of all the works on the power control 
due to the huge amount of papers published in the area. 

\vspace{0.2cm}

Starting from the differential equation form of the Sigmoid DPCA in 
\cite{UykanPhD01} and \cite{Uykan04} and relaxing the 
constraints on the positiveness and strict assumptions on the spectral radius of the link gain matrix, 
we establish a bridge from SgmDPCA to the Hopfield-like Neural Networks (NNs). 
The proposed approach yields a Sigmoid basis 
SIR-balancing NN which exhibits similar features as Hopfield NN does. 

Our investigations show that 
1) The proposed Sgm"SIR"NN includes both SgmDPCA algorithm and Hopfield NN as special cases. 
2) The Sgm''SIR''NN exhibits features which are generally attributed to Hofield-like recurrent NN.
3) Estalishing an analogy to the SgmDPCA, the proposed network as well as the Hopfield Network 
keeps the fictitious SIR at a target level. 


The paper is organized as follows:  The Hopfield Network and PC problem are investigated in the same framework in 
\ref{Section:model}.  Section \ref{SgmCIRNN} presents the proposed network. Simulation results are 
presented in Section \ref{Section:SimuResults} followed by Concluding Remarks in Section \ref{Section:CONCLUSIONS}.

\vspace{0.2cm}

\section{From Continuous-Time Sigmoid-Basis Power Control Algorithm to Hopfield-like NNs }
\label{Section:model}

We start with the standard definition of Signal-to-Interferende+Noise-Ratio (SIR) in a 
cellular radio system, in which $N$ mobiles share the same channel (e.g. \cite{Zander92a}, 
\cite{Zander92b}).  

\begin{equation} \label{eq:cir}
\gamma_i = \frac{ g_{ii} p_i}{ \nu_i + \sum_{j = 1, j \neq i}^{N} g_{ij} p_j },  \quad i=1, \dots, N
\end{equation}

where  $p_i$ is the transmission power of mobile $i$, $g_{ij}$ is
the link gain from mobile $j$ to base $i$ involving path loss,
shadowing, multi-path fading (as well as the spreading/processing gain in case of 
CDMA transmission \cite{Rappaport96}, etc), and $\nu_i$ is the
receiver noise at base station $i$.  

Without loss of generality, eq.(\ref{eq:cir}) considers the uplink case (from mobile to base) and 
assume that mobile $i$ is assigned to base $i$ at that instant. 
So, the aim of power control is to determine 
the transmit power for every mobile which keep 
its SIR (i.e., $\gamma_i$) at a target value $\gamma_i^{tgt}$.

Defining $H_{ij} = [ {\mathbf H} ]_{ij} = \gamma_{i}^{tgt} a_{ij}
/ a_{ii}$ and $H_{ii} = 0$ and $\eta_i = \gamma_{i}^{tgt} \nu_i /
a_{ii}$, one may obtain the eq. (\ref{eq:cir}) 
in matrix form as follows at the solution, i.e. when $\gamma_i = \gamma_i^{tgt}$.  

\begin{equation}\label{eq:cir_PC_matrix} 
 ({\mathbf I} - {\mathbf H}) {\bf p}  =  
{\mathbf \eta} 
\end{equation} 

where ${\bf p}$ is the power vector, ${\bf H}$ is the normalized
link gain matrix and 
${\mathbf \eta}$
is the noise vector. 

Considering the link gain matix ${\mathbf H} > 0$ (i.e., all entries are strictly positive), 
if the matrix $({\mathbf I} - {\mathbf H})$ is nonsingular and all its eigenvalues are strictly positive, then 
the positivity condition (i.e., the solution ${\mathbf p}^{*} > {\mathbf 0}$) is met 
because then $({\mathbf I} - {\mathbf H})^{-1} = {\mathbf I} + {\mathbf H} + {\mathbf H}^2 + 
\dots + {\mathbf H}^n + \dots $. 
Zander's work and all other works in the same line assumes that the spectral
radius of matrix ${\bf H}$ is smaller than 1. This is a sufficient condition for a unique positive solution. 
(For proof, see e.g. Theorem 3.7 in 
\cite{Varga62}).  The spectral radius of matrix ${\bf H}$ ($N \times N$) is defined as 
$
\max_i \{ |\lambda_i| \}_{i=1}^N$ where $\lambda_i$ are 
eigenvalues of ${\bf H}$.
In many works which is in the same line as Zander's, it's assumed that there is a network admission algorithm
which assures that the spectral radius of the 
normalized link gain matrix ${\mathbf H}$ is less than 1.

In the power control design, the link gain matrix and transmit power are all positive.
In what follows, we relax the positivity and radial spectral raduis conditions with the intend of 
having multiple equilibrium points to store a set of prototype vectors and then proceed to 
 Hopfield-like Neural Networks. 

Multiplying the eq. (\ref{eq:cir}) with $\frac{1}{\gamma_i^{tgt}}$ from both sides gives 
 
\begin{equation}\label{eq:cirAtgt}
\frac{\gamma_i}{\gamma_i^{tgt}} = \frac{ g_{ii} p_i}
{ \gamma_i^{tgt} \nu_i + \sum_{j = 1, j \neq i}^{N} (g_{ij} \gamma_i^{tgt}) p_j },   \quad i=1, \dots, N
\end{equation}

Let's define the following 
fictitious ``Signal to Interference Ratio (SIR)'' ($\theta_i$) by 
rewriting the eq.(\ref{eq:cirAtgt}) with neural network terminology: 

\begin{equation} \label{eq:cirA}
\frac{\theta_i}{\theta_{i}^{tgt}} = \frac{ a_{ii} x_i}{ b_i + \sum_{j = 1, j \neq i}^{N} w_{ij} x_j },  
	\quad i=1, \dots, N
\end{equation}

where  
$\theta_i$ is the defined fictitious ``SIR'', 
$x_i$ is the state of the $i$'th neuron, 
$a_{ii}$ is the feedback coefficient from its state to its input layer,  
$w_{ij}$ is the weight from the output of the $j$'th neuron to the input of the 
$j$'th neuron.  

From eq.(\ref{eq:cirA}), we define the following error signal $e_i$

\begin{equation} \label{eq:cirA_altrntf}
 e_{i} =  
-a_{ii} x_i + I_i,
\quad \quad  
\textrm{where} \quad I_i =  b_i + \sum_{j = 1, j \neq i}^{N} w_{ij} x_j,
\quad i=1, \dots, N
\end{equation}

The norm of the error signal vector ${\bf e}=[e_{1} \dots  e_{N}]^T$ can also be 
used as a performance index. 
In the second part of this work in \cite{Uykan08b}, 
$l1$ norm of ${\bf e}$ is examined in discrete time, and is shown to converge 
to ${\bf e} = {\bf 0}$ under some reasonable conditions both synchronously and asynchronously.

Prototype vectors are defined as those ${\mathbf x}$'s 
which make $\theta_i = \theta_{i}^{tgt} = 1, \quad i=1, \dots, N$ 
in eq.(\ref{eq:cirA}).  So, from 
eq.(\ref{eq:cirA}) and (\ref{eq:cirA_altrntf}), the prototype vectors 
make the error signal zero, i.e., $e_i=0, \quad i=1, \dots, N$ given that 
$x_i \neq 0$ and $I_i \neq 0$.

\begin{figure}[t!]
  \begin{center}
   \epsfxsize=18.0em    
\leavevmode\epsffile{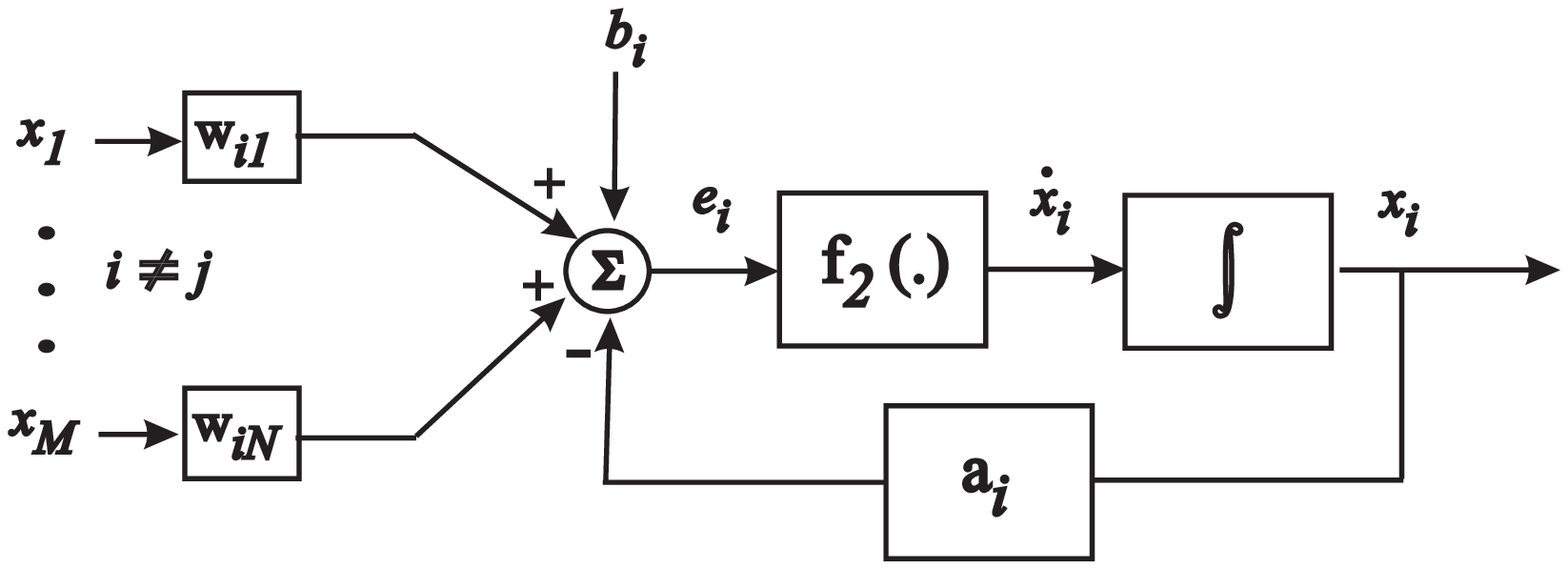}
   \vspace{-1em}        
  \end{center}
\begin{center} (a) \end{center}
  \begin{center}
   \epsfxsize=18.0em    
\leavevmode\epsffile{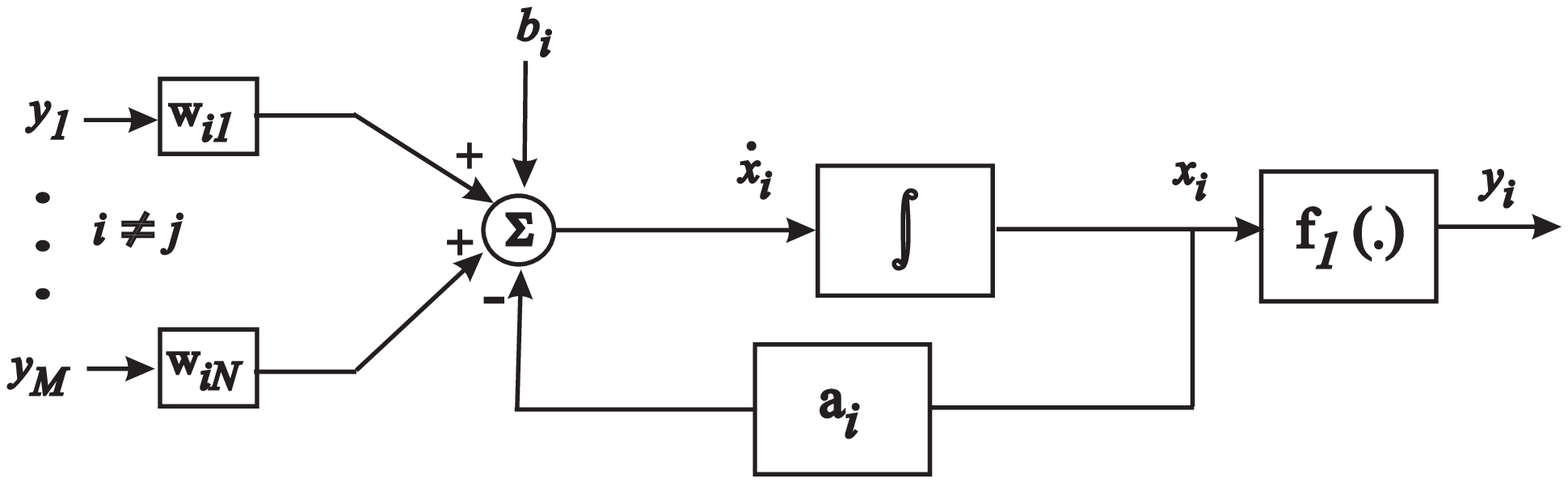}
   \vspace{-1em}        
  \end{center}
\begin{center} (b) \end{center}
  \begin{center}
   \epsfxsize=18.0em    
\leavevmode\epsffile{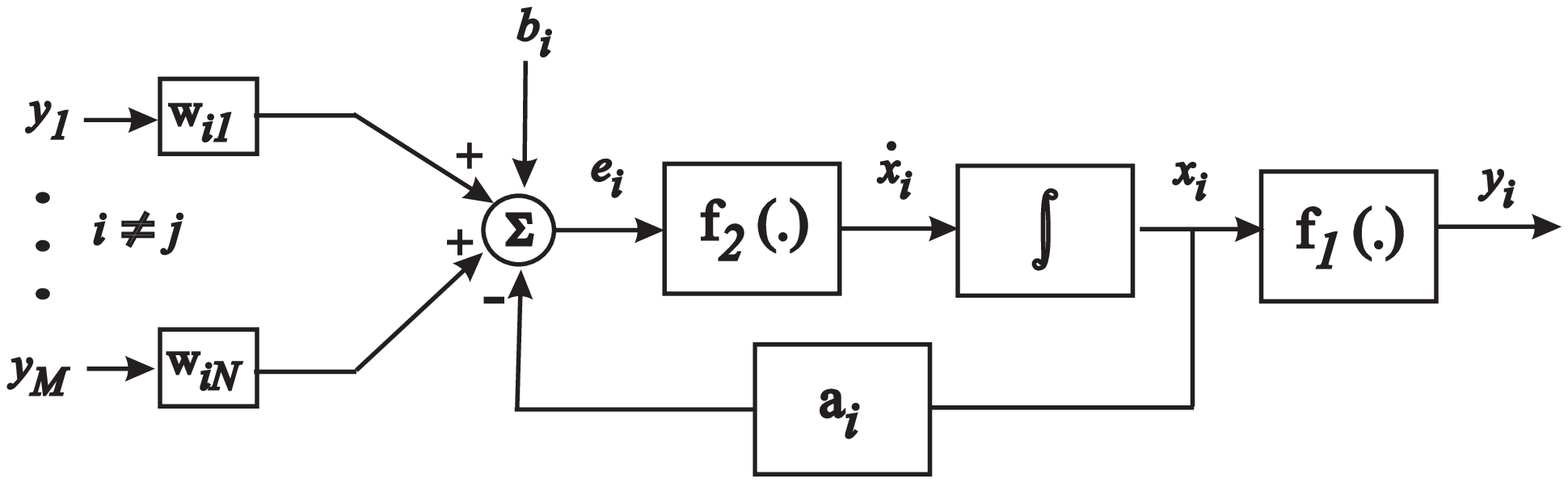}
   \vspace{-1em}        
  \end{center}
\begin{center} (c) \end{center}
 \caption
{ (a) SgmDPCA in \cite{UykanPhD01} and \cite{Uykan04}, (b) Hopfield Neural Network, 
(c) Proposed network, Sgm''SIR''NN, 
(Sgm``SIR''NN contains SgmDPCA and Hopfield NN as special cases). \label{fig:CIRHop} 
}
\end{figure}

The Fig. \ref{fig:CIRHop}.a 
shows the network corresponding to the differential equation form of the Sigmoid DPCA in \cite{UykanPhD01} 
and \cite{Uykan04}. The design problem in \cite{UykanPhD01} and \cite{Uykan04} is to device a 
stable dynamic Multiple Input Multiple Output (MIMO) 
network with the unique solution of ${\bf p}^{*} = ({\mathbf I} - {\mathbf H}) {\bf \eta}$ 
to be achieved as fast as possible. 
Note that the unique solution balances the SIRs of every mobile in its setting. 
Our paper extends the analysis to the multiple equilibrium points cases as follows: In this paper, we examine 

        a) the case where the states (transmit powers) are not bounded, (which is not a practical 
assumption in the power control algorithm). This case 
corresponds to Fig.\ref{fig:CIRHop}.a and is examined in the rest of this section.

        b) the cases where the states (transmit powers) are upper and lower bounded 
(inspired by the fact that the transmit power in the power control is lower and 
upper bounded in practise).  These cases are examined in Section \ref{SgmCIRNN}, 
which correspond to the networks in Fig.\ref{fig:CIRHop}.c 

In all these cases, we examine if the corresponding MIMO systems 
exhibit similar features 
as traditional continuous Hopfield Network does.

Writing the differential equation of the network in Fig.\ref{fig:CIRHop}.a in matrix form gives

\begin{equation} \label{eq:Diff_DPCA}
\dot{ {\mathbf x}} =  {\mathbf f}_1 \Big( -{\mathbf A}{\mathbf x} + {\mathbf W} \mathbf{x} + {\mathbf b} \Big)
\end{equation}

where $\dot{ {\mathbf x}}$ shows the derivative of ${\mathbf x}$ with respect to time, i.e., 
$\dot{ {\mathbf x}} = \frac{d{\mathbf x}}{dt}$, and $f_1(\cdot)$ represents the sigmoid function and   

\begin{equation} \label{eq:matA_W_b}
{\mathbf A} =
\left[
\begin{array}{c c c c}
a_{11}   &   0   & \ldots  &  0 \\
0     &   a_{22} & \ldots  &  0 \\
\vdots &      & \ddots  &  0 \\
0     &   0   & \ldots  &  a_{NN}
\end{array}
\right]
\quad \quad
{\mathbf W} =
\left[
\begin{array}{c c c c}
0  &   w_{12}   & \ldots  &  w_{1N} \\
w_{21}     &   0 & \ldots  &  w_{2N} \\
\vdots &     & \ddots  &  \vdots \\
w_{N1}    &   w_{N2}   & \ldots  &  0
\end{array}
\right]
\quad \quad
{\mathbf b} =
\left[
\begin{array}{c}
b_1 \\
b_2 \\
\vdots \\
b_N
\end{array}
\right]
\end{equation}

In eq.(\ref{eq:matA_W_b}), ${\mathbf A}$ shows the self-state-feedback matrix, ${\mathbf W}$ with zero 
diagonal shows the inter-neurons \
connection weight matrix, and ${\mathbf b}$ is a threshold vector. In (\ref{eq:Diff_DPCA}), the 
sigmoid function is $f_1(e) = 1 - \frac{1}{1 + exp(-\sigma_1 e)}$, where 
$\sigma_1>0$ is called the slope of $f_1(\cdot)$, which is equal to its derivative with respect to its argument 
at origin 0.

It's well known that desigining the weight matrix ${\mathbf W}$ as a symmetric one yields that all 
eigenvalues are real,
which we assume throughout the paper due to the simplicity and brevity of its analysis.



\vspace{0.2cm}
\emph{Proposition 1:} 
\vspace{0.2cm}

The network in eq. (\ref{eq:Diff_DPCA})


a) has unique equilibrium point and is globally stable if all the eigenvalues of matrix
$({\bf -A+W})$
are strictly negative, i.e. . $\lambda_i < 0, \quad , i=1,...,N$ where $\lambda_i$
are the eigenvalues of (${\bf -A+W}$).

b) is unstable in the sense that at least one $x_i(t)$ goes to infinity,
if at least one of the eigenvalues of matrix (${\bf -A + W}$) is strictly positive, i.e.
there exists a $\lambda_j > 0$. 

c) has infinite number of equlibrium points in the unbounded input space and finite number
of equilibrium points in a limited specified input space like a hybercube,
if the matrix $({\bf -A + W})$ is a negative semidefinite matrix (i.e., there exists only zero and negative
eigenvalues of $({\bf -A + W})$ ). So, by properly designing matrices ${\bf -A, W}$ for a specified input space
like a hybercube, a set of prototype vectors could be stored on the equilibrium points.


\begin{proof}

a) In what follows, we present a Lyapunov function for providing a sufficient condition for the stability 
of the network in (\ref{eq:Diff_DPCA}). 

A Lyapunov function candidate for (\ref{eq:Diff_DPCA}) is: 

\begin{equation} \label{eq:LyapDPCA}
V( \mathbf{x} ) = -\frac{1}{2} \mathbf{x}^T  ({\mathbf -A+W}) 
{\mathbf x} - \mathbf{ b }^T {\mathbf x}  
\end{equation}

where the matrix (${\bf -A + W}$) is a negative definite matrix 
which assures that the Lyapunov function in (\ref{eq:LyapDPCA}) is lower bounded. 

The derivative of the Lyapunov function with respect to time gives

\begin{equation} \label{eq:LyapDPCA_a}
\dot{V}( t ) = \frac{dV}{dt} = - \big( ({\mathbf -A+W}) {\mathbf x} + \mathbf{ b } \big)^T \bf{ \dot{x} }
\end{equation}


From eq.(\ref{eq:Diff_DPCA}) and (\ref{eq:LyapDPCA_a}) 

\begin{equation} \label{eq:LyapDPCA_b}
\dot{V}( t ) = - \mathbf{ f }^{-1}_1(\mathbf{ \dot{x} }^T) \mathbf{ \dot{x} }
\end{equation}

where $f^{-1}_1(\cdot)$ shows the inverse of the sigmoid function $f_1(\cdot)$.  
Since sigmoid function is an odd function and is zero if and only if its argument is zero, we obtain 
from (\ref{eq:LyapDPCA_b})


\begin{equation} \label{eq:LyapDPCA_c}
\dot{V}  
\left\{ 
\begin{array}{ll}
< 0 & \textrm{if and only if} \quad || \bf{ \dot{x} } || \neq \mathbf{ 0 }, \\
= 0 & \textrm{if and only if} \quad || \bf{ \dot{x} } || = \mathbf{ 0 } 
\end{array}
\right.
\end{equation}

It's well known that all the eigenvalues of a negative definite matrix is strictly negative and is 
nonsingular  (see e.g. \cite{Varga62}).  This implies that there is a unique solution for 
eq.(\ref{eq:Diff_DPCA}), i.e., $ {\bf x}^{eq} = ({\bf -A + W})^{-1}{\bf b} $.  This observation together with 
the Lyapunov function analysis above (eq. (\ref{eq:LyapDPCA}) and (\ref{eq:LyapDPCA_c}), 
proves part a.

b) The stability of the system eq.(\ref{eq:Diff_DPCA}) depends only 
on the matrix $({\bf -A + W})$, and 
not on the vector ${\bf b}$:   
Let the system be stable and let ${\mathbf x}^*$ be the equilibrium point of eq.(\ref{eq:Diff_DPCA}), i.e., 
$\Big( -{\mathbf A} + {\mathbf W} \Big) \mathbf{x}^* + {\mathbf b} = {\mathbf 0}$. 
Defining $\hat{ {\bf x}}(t) = {\bf x}(t) - {\mathbf x}^*$, and replacing it in eq.(\ref{eq:Diff_DPCA}) 
yields

\begin{equation} \label{eq:Diff_DPCA_bar}
\mathbf{ \dot{ \hat{ x } } } =  \mathbf{f}_1 
\Big( 
( -{\mathbf A} + {\mathbf W} ) \mathbf{ \hat{x} }
\Big)  
\end{equation}

As the state vector ${\bf x}(t)$ approaches to the ${\bf x}^*$ in eq.(\ref{eq:Diff_DPCA}), 
the $\hat{ {\mathbf x} }$ approaches to the origin ${\bf 0}$ in eq.(\ref{eq:Diff_DPCA_bar}). 
On the other hand, if the network in eq.(\ref{eq:Diff_DPCA}) is unstable, 
then the network (\ref{eq:Diff_DPCA_bar}) is also unstable, which 
shows that the stability of the system (\ref{eq:Diff_DPCA}) does not depend on the vector ${\bf b}$.
In brief, from stability point of view, 
it would be enough to examine only the matrix $({\bf -A + W})$, and not $({\bf b})$ in eq.(\ref{eq:Diff_DPCA}). 

If the matrix $({\bf -A + W})$ has a positive eigenvalue, then 
taking the initial state as the corresponding eigenvector, shown as $\mathbf{ x }^{eig}$, i.e.,   
$\mathbf{x}(t=0) = \mathbf{ x }^{eig}$ gives  

\begin{equation} \label{eq:Lemma1_partb} 
\mathbf{ \dot{ \hat{x} } } = \mathbf{ f }_1( \lambda_j \mathbf{ \hat{x} } ), \quad  
\mathbf{ \hat{x} }(0) = \mathbf{ x }^{eig}
\end{equation} 

where $\lambda_j$ shows the positive eigenvalue and $f_1(\cdot)$ is the sigmoid function.

Since $f(\cdot)$ is an odd function, eq.(\ref{eq:Lemma1_partb}) is unstable with $\lambda_j > 0$ 
simply because then the sign of the $d {\mathbf{ \hat{x} }} / dt$ is the same as the sign of 
${\mathbf{ \hat{x} }(t)}$ in eq.(\ref{eq:Lemma1_partb}), which proves part b.

c) Let's assume that the matrix ($\bf{-A + W}$) is a negative semidefinite matrix, and 
let's choose the same Lyapunov function candidate as in eq. (\ref{eq:LyapDPCA}), i.e.,  
$V( \mathbf{x} ) = -\frac{1}{2} \mathbf{x}^T  ({\mathbf -A+W}) {\mathbf x} - \mathbf{ b }^T {\mathbf x}$. 
Then, the Lyapunov function is lower bounded. Following the steps in part a above yields 

\begin{equation} \label{eq:LyapDPCA_partC}
\dot{V}( t ) = - \mathbf{ f }^{-1}(\mathbf{ \dot{x} }) \mathbf{ \dot{x} }
\end{equation}

which indicates that 

\begin{equation} \label{eq:LyapDPCA_partC}
\dot{V}
\left\{
\begin{array}{ll}
< 0 & \textrm{if and only if} \quad || \bf{ \dot{x} } || \neq \mathbf{ 0 }, \\
= 0 & \textrm{if and only if} \quad || \bf{ \dot{x} } || = \mathbf{ 0 }
\end{array}
\right.
\end{equation}

The equilibrium points of network in eq.(\ref{eq:Diff_DPCA}), which corresponds to 
Fig. \ref{fig:CIRHop}.a, satisfies the following linear equation:

\begin{equation} \label{eq:matSIRbalancing2}
\Big(
\left[
\begin{array}{c c c c}
a_{11}  &   0   & \ldots  &  0 \\
0     &   a_{22} & \ldots  &  0 \\
\vdots &      & \ddots  &  0 \\
0     &   0   & \ldots  &  a_{NN} 
\end{array}
\right] 
-
\left[
\begin{array}{c c c c}
0  &  w_{12}   & \ldots  &  w_{1N} \\
w_{21}     &   0 & \ldots  &  w_{2N} \\
\vdots &     & \ddots  &  \vdots \\
w_{N1}    &   w_{N2}   & \ldots  &  0
\end{array}
\right]
\Big)
\left[
\begin{array}{c}
x_1 \\
x_2 \\
\vdots \\
x_N
\end{array}
\right]
=
\left[
\begin{array}{c}
b_1 \\
b_2 \\
\vdots \\
b_N
\end{array}
\right] 
\end{equation}

On the other hand, writing the eq. (\ref{eq:cirA}) in matrix form gives 

\begin{equation} \label{eq:matSIRbalancing1}
\Big(
\left[
\begin{array}{c c c c}
a_{11} \frac{\theta_1^{tgt}}{\theta_1}   &   0   & \ldots  &  0 \\
0     &   a_{22} \frac{\theta_2^{tgt}}{\theta_2} & \ldots  &  0 \\
\vdots &      & \ddots  &  0 \\
0     &   0   & \ldots  &  a_{NN} \frac{\theta_N^{tgt}}{\theta_N}
\end{array}
\right] 
-
\left[
\begin{array}{c c c c}
 0  &  w_{12} & \ldots  &  w_{1N} \\
 w_{21}     &   0 & \ldots  &  w_{2N} \\
\vdots &     & \ddots  &  \vdots \\
w_{N1}    &  w_{N2}   & \ldots  &  0
\end{array}
\right]
\Big)
\left[
\begin{array}{c}
x_1 \\
x_2 \\
\vdots \\
x_N
\end{array}
\right]
=
\left[
\begin{array}{c}
 b_1 \\
 b_2 \\
\vdots \\
 b_N
\end{array}
\right] 
\end{equation}


Note that in eq.(\ref{eq:matSIRbalancing1}) the matrix 
$( \bf{W} )$ 
and 
$(\bf{b})$ 
are constant and only matrix 
$(\bf{A})$
varies as any of the $x_{i}$ changes because then all $\theta_{j}$'s \quad j=1, ..., N, change according 
to eq.(\ref{eq:cirA}).

Comparing eq.(\ref{eq:matSIRbalancing1}) and eq.(\ref{eq:matSIRbalancing2}), we choose $\theta_i^{tgt} = 1$ 
without loss of generality and for the sake of brevity.  
The equilibrium points of the network in (\ref{eq:matSIRbalancing2}) and those of the network of 
eq.(\ref{eq:matSIRbalancing1}) with $\theta_i^{tgt} = \theta_i = 1$ are equal.

Clearly, if we think of the whole input space, then there are infinite number
of equlibrium points due to the singularity of matrix ($\bf{-A + W}$) in (\ref{eq:matSIRbalancing2}).  
However, if we think of a certain input subspace assuming that 
the prototype vectors are in the corners of a hybercube as in
the case of many practical applications, and the initial states are within the 
hybercube, then there is a finite number of equlilibrium points within the 
hybercube. 

\end{proof}

In the anaysis above, we examined the network with no bound on the states, and show that the network is 
stable in the case of multiple equilibrium points in a bounded hybercube 
when there exists both zero and negative eigenvalues. 
However, in multiple equilibrium case, 
the simulation results suggest that 
its performance is quite poor especially when the dimension increases. 
Instead, in what follows, we investigate 
networks with lower and upper bounded 
states (similar as in the power control case where the maximum transmit power is 
lower and upper bounded in practice).  
Finally, we end up with a sigmoid based "SIR"-balancing network which exhibit similar properties 
as traditional Hopfield neural networks does.



\section{"SIR"-balancing Sigmoid Neural Network with bounded states \label{SgmCIRNN}}

In the power control formulation in section \ref{Section:model}, 
it was assumed that either there is no constraints 
on the maximum transmit power (which is an impractical assumption) or the existing unique solution is 
within the minimum and maximum transmit power constraints. 
In this section, by lower and upper bounding the system states of the proposed networks and relaxing 
the positivity condition, we present a Sigmoid based SIR-balancing networks which exhibits 
similar features as Hopfield NN does.



In practice, in power control the positive transmit power can not be arbitarily small and large.  
So, writing  eq.(\ref{eq:cir}) with the minimum and maximum power constraints gives 

\begin{equation} \label{eq:cirWithMaxTxp}
\gamma_i = \frac{ g_{ii} \max \{ p_{min}, \min \{p_{max}, p_i \} \} }
 	{ \nu_i + \sum_{j = 1, j \neq i}^{N} g_{ij} \max \{ p_{min}, \min \{p_{max}, p_j \} \} },  
	\quad i=1, \dots, N
\end{equation}

where $p_{min}$ and $p_{max}$ is the minimum and maximum transmit powers. The SIR model in 
(\ref{eq:cirWithMaxTxp}) can be further written in a more generalized equation as follows 

\begin{equation} \label{eq:cirFx}
\gamma_i = \frac{ g_{ii} y(p_i)}{ \nu_i + \sum_{j = 1, j \neq i}^{N} g_{ij} y(p_j) },  \quad i=1, \dots, N
\end{equation}

where $y(\cdot)$ represents the modeling of lower and upper bounding the transmit power and of any 
other effects e.g. power amplifier, etc. For example, 
$y(p_i) =  \max \{ p_{min}, \min \{p_{max}, p_i \} \}$ or corresponding piecewise linear function 
$y(p_i) = | p_i + p_{max} | - | p_i - p_{max} | $  yields eq.(\ref{eq:cirWithMaxTxp}).

By relaxing the positivity conditions in the power control problem in (\ref{eq:cirFx}) and using sigmoid  
as the bounding function to the states of the proposed network, we define the following fictitious "SIR":

\begin{equation} \label{eq:cirD}
\frac{\bar{\theta}_i}{\theta_{i}^{tgt}} =
      \frac{ a_{ii} f(x_i)}{ b_i + \sum_{j = 1, j \neq i}^{N} w_{ij} f(x_j) },  \quad i=1, \dots, N
\end{equation}

where $f(\cdot)$ represents the sigmoid function.  

Using sigmoid function in (\ref{eq:cirD}) will allow us 
to design Hopfield-like networks, which 
includes the traditional Hopfield Network as a special case, 
as will be seen in the following subsection.


Implementing the above-mentioned upper and lower bounds into the network in Fig.\ref{fig:CIRHop}.a 
(which was originally suggested in \cite{UykanPhD01} and \cite{Uykan04} for the power control problem) 
results in the following equation 


\begin{equation} \label{eq:DiffSgmNN_v1}
\dot{ {\bf x}} =  {\bf f}_1 \Big( -{\bf A}{\bf f}_2({\bf x}) + 
                            {\bf W}{\bf f}_2({\bf x}) + {\mathbf b} \Big)
\end{equation}


Let's define the following error signal for the network eq.(\ref{eq:DiffSgmNN_v1}) 

\begin{equation} \label{eq:DiffSgmNN_e_signal_elementwise}
e_i = -a_{ii} f_2(x_i) + I_i, \quad \quad 
\textrm{where} \quad I_i =  b_i + \sum_{j = 1, j \neq i}^{N} w_{ij} f_2(x_j), 
	\quad i=1, \dots, N
\end{equation}

Writing (\ref{eq:DiffSgmNN_e_signal_elementwise}) in matrix form gives  

\begin{equation} \label{eq:DiffSgmNN_e_signal}
{\bf e} = -{\bf A}{\bf f}_2({\bf x}) + 
                {\bf W}{\bf f}_2({\bf x}) + {\mathbf b}
\end{equation}
 
From eq. (\ref{eq:DiffSgmNN_v1}) and (\ref{eq:DiffSgmNN_e_signal}), $\dot{ {\bf x}} = {\bf f}_1({\bf e})$, 
which shows $\dot{ {\bf x}} = {\bf 0}$ if and only if ${\bf e} = {\bf 0}$ due to the 
chosen sigmoid function $f_1(\cdot)$.  
So, if $e_i = 0$ given that $x_i \neq 0$ and $I_i \neq 0$ in eq.(\ref{eq:DiffSgmNN_e_signal_elementwise}), 
then, from eq.(\ref{eq:cirD}) and (\ref{eq:DiffSgmNN_e_signal_elementwise}), 
$\bar{\theta}_i = \theta_i^{tgt} = 1$.

\vspace{0.2cm}
\emph{Proposition 2:} 
\vspace{0.2cm}

If $\mathbf{ W }$ is symmetric and $(\mathbf{ -A + W })$ is a negative semi-definite matrix, 
the network in (\ref{eq:DiffSgmNN_v1}) is stable, and the error vector 
${\bf e}$ in (\ref{eq:DiffSgmNN_e_signal}) goes to zero.  


\begin{proof} \label{pr:SgmSIRNN_Lyap_v1}

Let's examine the following Lyapunov function candidate for the network eq.(\ref{eq:DiffSgmNN_v1}) 

\begin{equation} \label{eq:LyapSgmCIR_v1}
V( \mathbf{x} ) = -\frac{1}{2} 
{\bf f}_2({\bf x}^T) ({\bf -A+W}) {\bf f}_2({\bf x}) - \mathbf{ b }^T {\mathbf f(x)}
\end{equation}

where $f_2(\cdot)$ represents the sigmoid function. Note that the Lyapunov function 
in (\ref{eq:LyapSgmCIR_v1}) is lower bounded for any $\bf{x}$ since matrix $({\bf -A+W})$ is 
negative semi-definite matrix. 
Next, we examine the derivative of the Lyapunov function 
with respect to time 

\begin{equation} \label{eq:LyapSgmCIR_v1_a}
\dot{V}( t ) =  
     		- \big( (\mathbf{-A+W}) {\bf f}_2({\bf x}) + {\bf b} \big)^T   \frac{d {\bf f}_2}{dt} 
\end{equation}

From eq.(\ref{eq:DiffSgmNN_v1}), 
$(\mathbf{-A+W}) {\bf f}_2({\bf x}) + {\bf b} = {\bf f}_1^{-1}( {\bf \dot{x}} )$.  Using that in 
(\ref{eq:LyapSgmCIR_v1_a}) gives 

\begin{equation} \label{eq:LyapSgmCIR_v1_a1}
\dot{V}( t ) = 
- \left[
\begin{array}{c c c}
f_1^{-1}(\dot{x}_1) &  \ldots  &  f_1^{-1}(\dot{x}_N)  
\end{array}
\right]
\left[
\begin{array}{c}
\frac{df_2}{dx_1} \dot{x}_1 \\
\vdots \\
\frac{df_2}{dx_N} \dot{x}_N 
\end{array}
\right]
\end{equation}

and

\begin{equation} \label{eq:LyapSgmCIR_v1_b}
\dot{ V } = - \sum_{j = 1}^{N} \frac{df_2}{dx_i} f_{1}^{-1}(\dot{x}_i) \dot{x}_i
\end{equation}

Since the inverse function of the sigmoid function 
is an odd function, $f_1^{-1}(\cdot)$, and is zero if and only if 
its argument is zero, 
and $\frac{df_2}{dx_i} > 0$ in the work regime, we obtain

\begin{equation} \label{eq:LyapSgmCIR_v1_c}
\dot{V}
\left\{
\begin{array}{ll}
< 0 & \textrm{if and only if} \quad || \bf{ \dot{x} } || \neq \mathbf{ 0 }, \\
= 0 & \textrm{if and only if} \quad || \bf{ \dot{x} } || = \mathbf{ 0 }
\end{array}
\right.
\end{equation}

From eq.(\ref{eq:DiffSgmNN_v1}) and eq.(\ref{eq:DiffSgmNN_e_signal}), $\dot{{\bf x}} = {\bf f}_1({\bf e})$. 
Since $f_1(\cdot)$ is an increasing odd function and is zero only at origin 0, 
$\dot{{\bf x}} = {\bf 0}$ if and only if ${\bf f}_1({\bf e}) = {\bf 0}$. This observation together with 
eq.(\ref{eq:LyapSgmCIR_v1_c}) completes the proof. 

\end{proof}

\vspace{0.2cm}


In what follows, we examine the evolution of an energy function that gives an insight into the 
evolution of the error vector that is defined as the argument of the function 
${\mathbf f}_1 (\cdot)$ in a further generalized network.  As an attempt to better examine the roles 
of the diagonal matrix ${\mathbf A}$ and matrix ${\mathbf W}$ onto 
the network dynamic behaviour, let us consider possibly a different function for matrix ${\mathbf A}$, 
denoted as $f_3(\cdot)$, in eq.(\ref{eq:DiffSgmNN_v1}) as follows: 

\begin{equation} \label{eq:Appn_DiffSgmNN}
\dot{ {\mathbf x}} =  {\mathbf f}_1 \Big( -{\mathbf A}{\bf f}_3({\bf x}) + 
			{\bf W} {\bf f}_2({\bf x}) + {\mathbf b} \Big)
\end{equation}

where $f_1(\cdot)$, $f_2(\cdot)$ are 
sigmoid functions with possibly different (positive) slopes, and 
$f_3(\cdot)$ indicates the function implemented to ${\mathbf A}$.

So, the corresponding error vector is defined as  

\begin{equation} \label{eq:Appn_e}
{\bf e} = -{\bf A}{\bf f}_3({\bf x}) + 
                {\bf W}{\bf f}_2({\bf x}) + {\mathbf b}
\end{equation}

Let's define the energy function for the error vector in (\ref{eq:Appn_e}) as follows   

\begin{equation} \label{eq:Appn_Lyap}
V(t) = \frac{1}{2} {\bf f}_1({\bf e}^T) {\bf f}_1({\bf e}) 
\end{equation}

The derivative of the energy function with respect to time gives 

\begin{eqnarray} \label{eq:Appn_Lyap_der}
\dot{ V }  &  =  &  
{\bf f}_1({\bf e}^T) \frac{ d{\bf f}_1 ({\bf e}) }{dt} \\
   	&   =   &  
\left[
\begin{array}{c c c c}
\dot{x}_1   &  \dot{x}_2  &  \ldots  &  \dot{x}_N  
\end{array}
\right]
\left[
\begin{array}{c}
\frac{\delta f_1}{\delta e_1} 
 [ -a_{11} \frac{\delta f_3}{\delta x_1} + \sum_{j = 1, j \neq 1}^{N} w_{1j} \frac{\delta f_2}{\delta x_j} \dot{x}_j ] \\
\frac{\delta f_1}{\delta e_2} 
 [ -a_{22} \frac{\delta f_3}{\delta x_2} + \sum_{j = 1, j \neq 2}^{N} w_{2j} \frac{\delta f_2}{\delta x_j} \dot{x}_j ] \\
\vdots \\ 
\frac{\delta f_1}{\delta e_N} 
 [ -a_{NN} \frac{\delta f_3}{\delta x_N} + \sum_{j = 1, j \neq 2}^{N} w_{Nj} \frac{\delta f_2}{\delta x_j} \dot{x}_j ] \\
\end{array}
\right]  \\
	&  =  &  \dot{ {\mathbf x} }^T {\mathbf J} \dot{ {\mathbf x} }
\end{eqnarray}

where matrix ${\mathbf J}$ is equal to 

\begin{equation} \label{eq:Appn_JacobianMtrx}
{\mathbf J} = 
\left[
\begin{array}{c c c c}
-a_{11} \frac{\delta f_3}{\delta x_1} \frac{\delta f_1}{\delta e_1}  &   
	w_{12} \frac{\delta f_2}{\delta x_2} \frac{\delta f_1}{\delta e_1}   & 
		\ldots  &  
			w_{1N} \frac{\delta f_2}{\delta x_N} \frac{\delta f_1}{\delta e_1} \\
w_{21} \frac{\delta f_2}{\delta x_1} \frac{\delta f_1}{\delta e_2}  &   
	-a_{22} \frac{\delta f_3}{\delta x_2} \frac{\delta f_1}{\delta e_2}  & 
		\ldots  &  
			w_{2N} \frac{\delta f_2}{\delta x_N} \frac{\delta f_1}{\delta e_2} \\
\vdots &      & \vdots  &  \vdots \\
w_{N1} \frac{\delta f_2}{\delta x_1} \frac{\delta f_1}{\delta e_N}  &   
	a_{N2} \frac{\delta f_2}{\delta x_2} \frac{\delta f_1}{\delta e_N}  & 
		\ldots  &  
			-a_{NN} \frac{\delta f_3}{\delta x_N} \frac{\delta f_1}{\delta e_N}
\end{array}
\right]
\end{equation}

where $a_{ii} >0$.  If matrix ${\mathbf J}$ in eq.(\ref{eq:Appn_JacobianMtrx}) 
is negative definite, then the error vector in (\ref{eq:Appn_e}) goes to zero due to 
eq.(\ref{eq:Appn_Lyap}) and (\ref{eq:Appn_Lyap_der}).  

From eq.(\ref{eq:Appn_JacobianMtrx}), we observe that 

a) $\frac{\delta f_1}{\delta e_i}$ is seen in every element of raw $i$. So, it has no effect 
on the negative definiteness of matrix ${\mathbf J}$. 

b) $\frac{\delta f_2}{\delta x_i}$ takes place in all non-diagonal elements, and not 
in the diagonal elements. 
From the characteristics of the derivative of the sigmoid function, 
$\frac{\delta f_2}{\delta x_i} \approx 0$  whenever $x_i$ is close to or in the saturation regime. 
This assures that the energy function decreases because $\dot{ V } <0$ in (\ref{eq:Appn_Lyap_der}), 
whenever all the $x_i$'s are in a saturation regime, 
provided that $\frac{\delta f_3}{\delta x_i} >> 0$.  

c) $\frac{\delta f_3}{\delta x_i}$ is seen on the diagonal elements only. 
So, taking the observations in a and b 
into account, if $f_3(\cdot)$ is chosen a unity function, i.e., $f_3(x_i)=x_i$, then 
the $\frac{\delta f_3}{\delta x_i} >> 0$ for the saturation regime, which assures that 
$\dot{ V } <0$ for any $\bf{x}$ in the saturation regime.


In the following subsection, we show that choosing $f_3(\cdot)$ as a unity function 
results a network which exhibits similar features as traditional 
Hopfield Network does.





\vspace{0.2cm}




\subsection{"SIR"-balancing Sigmoid Neural Network  \label{SgmCIRNNFx_v2}}

From the observations a, b and c above, we choose $f_3(\cdot)$ a unity function, i.e., $f_3(x_i)=x_i$. 
In a smililar way as in (\ref{eq:cirD}), we define the following "SIR":

\begin{equation} \label{eq:cirC}
\frac{\hat{\theta}_i}{\theta_{i}^{tgt}} = 
       \frac{ a_{ii} x_i}{ b_i + \sum_{j = 1, j \neq i}^{N} w_{ij} f_2(x_j) },  \quad i=1, \dots, N
\end{equation}

where $f_2(\cdot)$ is a sigmoid function and all other parameters are defined as in eq.(\ref{eq:cirD}). 
So, the only difference 
is that the self-state-feedback is directly from its state  
in eq.(\ref{eq:cirC}) as contrast to its function output in eq.(\ref{eq:cirD}). 
The resulting network is presented in Fig.\ref{fig:CIRHop}.c. 

From Fig.\ref{fig:CIRHop}.c, 

\begin{equation} \label{eq:DiffSgmNN_d}
\dot{ {\mathbf x}} =  {\mathbf f}_1 \Big( -{\mathbf A} {\bf x} + 
			{\bf W} {\bf f}_2({\bf x}) + {\mathbf b} \Big)
\end{equation}

We will call the network in eq.(\ref{eq:DiffSgmNN_d}) 
as Sgm"SIR"NN (Sigmoid ``SIR''-balancing neural network). 

Let's define the following error signal for the network eq.(\ref{eq:DiffSgmNN_d}) 

\begin{equation} \label{eq:DiffSgmNN_e_signal_elementwise_v2}
e_i = -a_{ii} x_i + I_i, \quad \quad 
\textrm{where} \quad I_i =  b_i \sum_{j = 1, j \neq i}^{N} w_{ij} f_2(x_j), 
	\quad i=1, \dots, N
\end{equation}

Writing (\ref{eq:DiffSgmNN_e_signal_elementwise_v2}) in matrix form gives 

\begin{equation} \label{eq:DiffSgmNN_e_signal_v2}
{\bf e} = -{\bf A}{\bf x} + 
                {\bf W}{\bf f}_2({\bf x}) + {\mathbf b}
\end{equation}

From eq. (\ref{eq:DiffSgmNN_d}) and (\ref{eq:DiffSgmNN_e_signal_v2}), 
$\dot{ {\bf x}} = {\bf f}_1 ( {\bf e} ) $.  
So, if $e_i = 0$ given that $x_i \neq 0$ and $I_i \neq 0$, then, 
from eq.(\ref{eq:cirC}) and (\ref{eq:DiffSgmNN_e_signal_elementwise_v2}), 
$\hat{\theta}_i = \theta_i^{tgt} = 1$.

The Sgm''SIR''NN in eq.(\ref{eq:DiffSgmNN_d}) includes both the sigmoid power control 
algorithm in \cite{Uykan04}, which is shown in eq.(\ref{eq:Diff_DPCA}),  
and the traditional Hopfield NN as 
special cases. (Taking $\mathbf{f}_2(.)$ as unity function 
results in sigmoid power control algorithm and 
taking $\mathbf{f}_1(.)$ as unity function results in Hopfield Network).

\vspace{0.2cm}
\emph{Proposition 3:} 
\vspace{0.2cm}

The Sgm"SIR"NN in eq.(\ref{eq:DiffSgmNN_d}) converges to $\mathbf{e} = \mathbf{0}$ if 

\begin{equation} \label{eq:lemma1_if}
| a_{jj} | \geq  \sum_{i=1, (i \neq j)}^{N}  |w_{ij}|
\end{equation}


\begin{proof} \label{pr:SgmSIRNN_DiagDominant}

Let's choose the Lyapunov function candidate as in (\ref{eq:Appn_Lyap}) for the network eq.(\ref{eq:DiffSgmNN_d}) 
as follows 

\begin{equation} \label{eq:DiagDominant_Lyap}
V(t) = \frac{1}{2} {\bf f}_1({\bf e}^T) {\bf f}_1({\bf e}) 
\end{equation}

where ${\bf e}$ is defined in eq.(\ref{eq:DiffSgmNN_e_signal_elementwise_v2}). Following the steps 
from (\ref{eq:Appn_Lyap_der}) to (\ref{eq:Appn_JacobianMtrx}), and taking the observations a to c 
into account, it's seen that if 

\begin{equation} \label{eq:lemma1_if_tekrar}
| a_{jj} | \geq | \sum_{i=1, (i \neq j)}^{N}  |w_{ij}|
\end{equation}

then 

\begin{equation} \label{eq:LyapDPCA_DiagDominant}
\dot{V}  
\left\{ 
\begin{array}{ll}
< 0 & \textrm{if and only if} \quad || {\bf f}_1({\bf e}) || \neq \mathbf{ 0 }, \\
= 0 & \textrm{if and only if} \quad || {\bf f}_1({\bf e}) || = \mathbf{ 0 } 
\end{array}
\right.
\end{equation}

which completes the proof, since $\dot{ {\bf x}} = {\bf f}_1 ( {\bf e} )$ from 
eq. (\ref{eq:DiffSgmNN_d}) and (\ref{eq:DiffSgmNN_e_signal_v2}). 

\end{proof}

The analysis above is for a diagonally dominant matrix case. In what follows, we 
examine for any positive $\mathbf{A}$ and symmetric $\mathbf{W}$ case. 

\vspace{0.2cm}
\emph{Proposition 4:} 
\vspace{0.2cm}

For the Sgm"SIR"NN in eq.(\ref{eq:DiffSgmNN_d}) with symmetric $\mathbf{W}$, and positive diagonal 
matrix $\mathbf{A}$, the defined error vector in (\ref{eq:DiffSgmNN_e_signal_elementwise_v2}) 
as well as the state vector $\mathbf{x}(t)$ stay within a bounded space for any time $t$.

\begin{proof} \label{pr:SgmSIRNN_AlmostLyapunov}

Let's choose the same energy function in eq.(\ref{eq:DiagDominant_Lyap}). 
Following the steps from eq.(\ref{eq:Appn_Lyap_der}) to (\ref{eq:Appn_JacobianMtrx}), and 
taking the observations a, b and c into account, we see that: The 
$\frac{\delta f_2}{\delta x_i} \approx 0$  for any $x_i$ which is close to or in the saturation regime
due to the characteristics of the derivative of the sigmoid function. 
This assures that the energy function decreases in any saturation regime
because $\dot{ V } <0$ in (\ref{eq:Appn_Lyap_der})  
for any $\mathbf{x}(t)$ in saturation. Therefore, the norm of (\ref{eq:DiffSgmNN_e_signal_v2})
do not go to infinity and stay within a bounded space.  This implies that the 
$\mathbf{x}(t)$ also stays within a bounded space from (\ref{eq:DiffSgmNN_e_signal_v2}). 

\end{proof}

We show in Proposition 4 that the $\mathbf{x}(t)$ stay within a bounded space for any time $t$. Since 
the number of all possible in-saturation state combinations is finite, which is equal to $2^N$, 
one may expect that either the network might show a sort of
an oscilatory behaviour which would never converge to $\mathbf{e}=\mathbf{0}$ or 
the network will eventually converge to an equilibrium point within a finite 
time, say $T_d$, satisfying $\mathbf{e}(t \geq T_d) = \mathbf{0}$.  
In what follows we show that 
the latter one is correct, i.e., the network converges to $\mathbf{e}=\mathbf{0}$.

\vspace{0.2cm}
\emph{Proposition 5:} 
\vspace{0.2cm}

The Sgm''SIR''NN in eq.(\ref{eq:DiffSgmNN_d}) with symmetric matrix {\bf W} and 
positive diagonal matrix $\mathbf{A}$ minimizes 
the following energy function (Lyapunov function) in (\ref{eq:LyapunovHop}) and exhibits similar features as 
traditional Hopfield Network does. 
The error vector ${\bf e}$ in (\ref{eq:DiffSgmNN_e_signal_v2}) goes to zero. 

\begin{equation} \label{eq:LyapunovHop} 
V( \mathbf{x} ) = -\frac{1}{2} \mathbf{f(x)}^T \mathbf{W f(x) } + \sum_{j = 1}^{N} \{ \int_{0}^{y_i} 
f_2^{-1}(u)du \} - \mathbf{b}^T \mathbf{f(x)} 
\end{equation} 

where $y_i = f_2(x_i)$ and $f_2^{-1}(y_i)$ represents the inverse of the sigmoid function. 
\footnote{
It's shown by Hopfield et al. in \cite{Hopfield85} that Hopfield network minimizes 
the energy (Lyapunov) function in eq.(\ref{eq:LyapunovHop}) 


and its derivative with respect to time is obtained in \cite{Hopfield85} 
as follows 

\begin{equation} \label{eq:LyapHopDerv} 
\frac{ dV }{dt} = - \sum_{j = 1}^{N} \frac{d f}{d x_i} ( \frac{d x_i}{dt} )^2 \leq 0  \nonumber
\end{equation} 
}


\begin{proof} \label{pr:SgmSIRNN_Lyap1}

Taking the derivative of the energy function (Lyapunov function candidate) in (\ref{eq:LyapunovHop}) 
with respect to time gives 

\begin{eqnarray} \label{eq:LyapCIRNN}
\dot{ V }(t)  &  =  &  
- {\bf f}_2({\bf x}) {\mathbf W} \frac{ d{\mathbf f}_2 }{dt}  
+ \sum_{j = 1}^{N} \frac{d}{dt}  \{ \int_{0}^{y_i} f^{-1}(u)du \} 
- {\mathbf b }^{T} \frac{ d{\mathbf f}_2 }{dt}   \\
 	&  =  & - {\bf f}_2({\bf x}) {\mathbf W} \frac{ d{\mathbf f}_2 }{dt}  
+ \sum_{j = 1}^{N} \frac{d}{df_2}  \{ \int_{0}^{y_i} f^{-1}(u)du \} \frac{df_2}{dt} 
- {\mathbf b }^{T} \frac{ d{\mathbf f}_2 }{dt}   
\end{eqnarray}

Since $\frac{d}{df_2}  \{ \int_{0}^{y_i} f^{-1}(u)du \} = f^{-1}(y_i)$, we obtain 

\begin{eqnarray}
\label{eq:LyapunovCIRNN_1}
\dot{ V }(t)  
 	&  =  &  - {\bf f}_2({\bf x}) {\mathbf W} \frac{ d{\mathbf f}_2 }{dt}  
+ \sum_{j = 1}^{N} a_{ii} f^{-1}(y_i) - {\mathbf b }^{T} \frac{ d{\mathbf f}_2 }{dt}  \\
 	&  =  &  [ - \mathbf{f(x)^T} {\mathbf W}^T + 
	{\bf x}^T {\bf A}^T  - {\mathbf b }^T ] \frac{ d{\mathbf f}_2 }{dt}  \label{eq:LyapunovCIRNN_2}
\end{eqnarray}

In (\ref{eq:LyapunovCIRNN_1}), $ f_2^{-1}(y_i) = x_i$ is used.   
From eq.(\ref{eq:DiffSgmNN_d}) and (\ref{eq:LyapunovCIRNN_2})

\begin{equation} \label{eq:LyapCIRNN1b}
\dot{ V }(t) = -[ {\mathbf f}_1 ( \dot{ {\mathbf x}} ) ]^T \frac{ d{\mathbf f}_2 }{dt}  
\end{equation}

and finally 

\begin{equation} \label{eq:LyapCIRNN1c}
\dot{ V } = - \sum_{j = 1}^{N} \frac{df_2}{dx_i}  f_1^{-1}(\dot{x}_i) \dot{x}_i
\end{equation}

Since sigmoid $f_2(.)$ is an increasing odd function, and $\frac{df_2}{dx_i} > 0$ in the work regime, 

\begin{equation} \label{eq:LyapSgmNN_v2}
\dot{V}
\left\{
\begin{array}{ll}
< 0 & \textrm{if and only if} \quad || \bf{ \dot{x} } || \neq \mathbf{ 0 }, \\
= 0 & \textrm{if and only if} \quad || \bf{ \dot{x} } || = \mathbf{ 0 }
\end{array}
\right.
\end{equation}

Eq. (\ref{eq:LyapSgmNN_v2}) shows that the Lyapunov function decreases at all points other 
than the equilibrium points and does not change only at the equilibrium points
where the error vector ${\bf e}$ in (\ref{eq:DiffSgmNN_e_signal_v2}) is zero vector, 
because $\bf{ \dot{x} } = \bf{ f }_1 (\bf{ e })$ where $f_1(\cdot)$ is sigmoid function. 
This completes the proof.

\end{proof}

\emph{Corollary 1:} 
\vspace{0.2cm}

Note that in the Lyapunov function analysis above, there is no assumption on the eigenvalues of the matrix 
${\mathbf W }$, and the diagonal matrix ${\mathbf A }$.  The only assumption is that ${\mathbf W }$ is symmetric 
and ${\mathbf A }$ is positive. 

So, from the Lyapunov analysis above 
for symmetric $\mathbf{ W }$ and positive $\mathbf{ A }$,  
we conclude that  

\begin{enumerate}

\item 
The Sgm''SIR''NN does not show oscilatory behaviour. (This is because 
the energy function of the Sgm''SIR''NN-v1 decreases at all points other 
than the equilibrium points and does not change only at the equilibrium points).



\item 
The states for any initial condition converge to one of the 
equilibrium points depending on the initial contition.  
If in the converged eqiulibrium point, $\theta_i = \theta_i^{tgt}=1$, then 
it corresponds to a prototype vector.

\item 
All equilibrium points are potential attractors. Equavalantly, all the attractors are stationary points 
of $V({\mathbf x})$.

\end{enumerate}



Note that the features above are also attributed to the traditional 
Hopfield NN in \cite{Hopfield85}. 
Taking $\mathbf{f}_1(.)$ as unity function results in Hopfield NN. 


From (\ref{eq:DiffSgmNN_d}), the equilibrium points of the Sgm"SIR"NN are the same as 
those of corresponding Hopfield NN. 
However, the basins of the attractors of the proposed Sgm"SIR"NN are, in general, different 
than those of the Hopfield NN, as will be seen from the simulation results in section 
\ref{Section:SimuResults}.



There are various ways for determining the weight matrix of the Hopfield Networks: Gradient-descent 
supervised learning (e.g. \cite{Haykin99}), 
solving linear inequalities (e.g. \cite{Berg96}, \cite{Harrer91} among others), 
Hebb learning rule \cite{Hebb49}, \cite{Muezzinoglu04} etc. 
How to design CINR-SgmNN 
is out of the scope of this paper. The methods used for traditional Hopfied NN can also be used for the 
Sgm``CIR''NN. 

As far as the simulation results in section \ref{Section:SimuResults} 
are concerned, for the sake of simplicity and brevity, we assume that the desired 
prototype vectors are orthogonal and we use the following design procedure 
for matrices $\mathbf{ A }$, $\mathbf{ W }$ and $\mathbf{ b }$, which is based on 
Hebb learning (\cite{Hebb49}): 


\vspace{0.2cm}
\emph{ Outer products based network design: }
\vspace{0.2cm}

Let's assume that $L$ desired orthogonal prototype vectors, $\{ \mathbf{ d }_s \}_{s=1}^{L}$, 
are chosen form  $(-1, +1)^N$.


Step 1: Calculate the sum of outer products of the prototype vectors (Hebb Rule, \cite{Hebb49})

\begin{eqnarray} \label{eq:HebbQd}
\mathbf{ Q } = \sum_{s=1}^{L} \mathbf{ d }_s  \mathbf{ d }_s^T
\end{eqnarray}

Step 2: Determine the diagonal matrix $\bf{A}$ and $\bf{W}$ as follows:

\begin{equation} \label{eq:AfromHebb}
a_{ij} = 
\left\{
\begin{array}{ll}
q_{ii} + \rho & \textrm{if} \quad i = j, \\
0 & \textrm{if} \quad i \neq j
\end{array}
\right.  \quad \quad i,j=1, \dots, N
\end{equation}

where $\rho$ is a real number and   

\begin{equation} \label{eq:WfromHebb}
w_{ij} =
\left\{
\begin{array}{ll}
0 & \textrm{if} \quad i = j, \\
q_{ij} & \textrm{if} \quad i \neq j
\end{array}
\right.   \quad \quad i,j=1, \dots, N
\end{equation}

where $q_{ij}$ shows the entries of matrix $\mathbf{ Q }$, $N$ is the dimension of the vector 
$\mathbf{ x }$ and $L$ is the number of the prototype vectors ($N > L > 0$). In eq.(\ref{eq:AfromHebb}), 
$q_{ii} = L$ from (\ref{eq:HebbQd}) since $\{ \mathbf{ d }_s \}$ is from $(-1, +1)^N$. 
It's observed that $\rho=0$ gives relatively good performance, however, 
by examining the nonlinear state equations in  eq.(\ref{eq:DiffSgmNN_d}), 
it can be seen that the proposed network Sgm"SIR"NN contains the prototype vectors 
at their equilibrium points for a relatively large interval of $\rho$ 
thanks to the bounding effect of the sigmoid function.

Another choice of $\rho$ in (\ref{eq:AfromHebb}) is $\rho = N-2L$ which yields $a_{ii} = N-L$.  
In what follows we show that this choice also assures that $\{ \mathbf{ d }_j  \}_{j=1}^L$ 
are the equilibrium points of the networks.



From (\ref{eq:HebbQd})-(\ref{eq:WfromHebb})

\begin{equation} \label{eq:HebbDPCA}
[ - \mathbf{ A } + \mathbf{ W } ] = -(N-L) \mathbf{ I } + 
    \sum_{s=1}^{L} \mathbf{ d }_s  \mathbf{ d }_s^T - L  \mathbf{ I } 
\end{equation}

where $\mathbf{ I }$ represents the identity matrix. 

Since $\mathbf{ d }_s \in (-1,+1)^N$, then  $|| \mathbf{ d }_s ||_2^2 = N$.  
Using (\ref{eq:HebbDPCA}) and the orthogonality 
properties of the set $\{ \mathbf{ d }_s  \}_{s=1}^L$ gives 

\begin{equation} \label{eq:HebbDPCA_a}
[ - \mathbf{ A } + \mathbf{ W } ] \mathbf{ d }_s  = -(N-L) \mathbf{ d }_s  + (N-L) \mathbf{ d }_s  = \mathbf{ 0 }
\end{equation} 

So, the prototype vectors $\{ \mathbf{ d }_j \}_{j=1}^L$ correspond to equilibrium points. 



\vspace{0.2cm}

\section{Simulation Results  \label{Section:SimuResults} }

In this section, we present two examples, one with 8 neurons and one with 16 neurons. 
The weight matrices are designed by the outer products-based design above.
Traditional Hopfield network is used a reference network.  
The continuous Hopfield Network \cite{Hopfield85} is 

\begin{equation} \label{eq:continuousHopfield}
\dot{ {\bf x}} =  -{\bf A}{\bf f}_2({\bf x}) +
                            {\bf W}{\bf f}_2({\bf x}) + {\mathbf b}
\end{equation}

where ${\bf A}$, ${\bf W}$, ${\bf b}$ and $f_2(\cdot)$ is defined as in eq.(\ref{eq:DiffSgmNN_d}).

\vspace{0.2cm}
\emph{Example 1:}
\vspace{0.2cm}
 
In this example, there are 8 neurons.  The desired prototype vectors are 

\begin{equation} \label{eq:ex1_D}
{\mathbf D} =
\left[
\begin{array}{c c c c c c c c}
1   &   1   & 1  &  1  & -1  &  -1  & -1  & -1  \\
1   &   1   & -1  &  -1  & 1  &  1  & -1  & -1  \\
1   &   -1  &  1  &  -1  & 1  &  -1  & 1  & -1 
\end{array}
\right]
\end{equation}

The weight matrices $\bf{ A }$ and $\bf{ W }$, and the threshold vector $\bf{ b }$ 
are obtained as follows 
by using the outer-products-based design presented in section \ref{SgmCIRNNFx_v2} and 
the slopes of sigmoid functions $f_1(\cdot)$ and $f_2(\cdot)$ are set to $\sigma_1=10$ and 
$\sigma_2=2$ respectively, and $\rho$ is chosen as -1.

\begin{equation} \label{eq:matA_W_b_ex1}
{\mathbf A} = 2{\mathbf I},  
\quad \quad
{\mathbf W} =
\left[
\begin{array}{c c c c c c c c}
0   &   1   &  1  &  -1  &  1  &  -1 & -1  & -3 \\
1   &   0   & -1  &   1  & -1  &  1  & -3  & -1 \\
1   &   -1  &  0  &   1  & -1  &  -3 & 1  & -1 \\
-1  &   1   & 1   &  0   & -3  &  -1 & -1 &  1 \\
1   &   -1  & -1  &  -3  & 0  &  1   &  1  & -1 \\
-1  &   1   &  -3 &  -1  & 1  &  0   & -1  & 1 \\
-1  &   -3  & 1   &  -1  &  1 &  -1  &  0  & 1 \\
-3  &   -1  & -1  &   1  & -1 &  1  &  1  & 0 
\end{array}
\right],  
\quad \quad
{\mathbf b} = {\mathbf 0}
\end{equation}

where ${\mathbf I}$ shows the identity matrix of dimension $N$ by $N$.

\begin{figure}[htbp]
  \begin{center}
   \epsfxsize=30.0em    
\leavevmode\epsffile{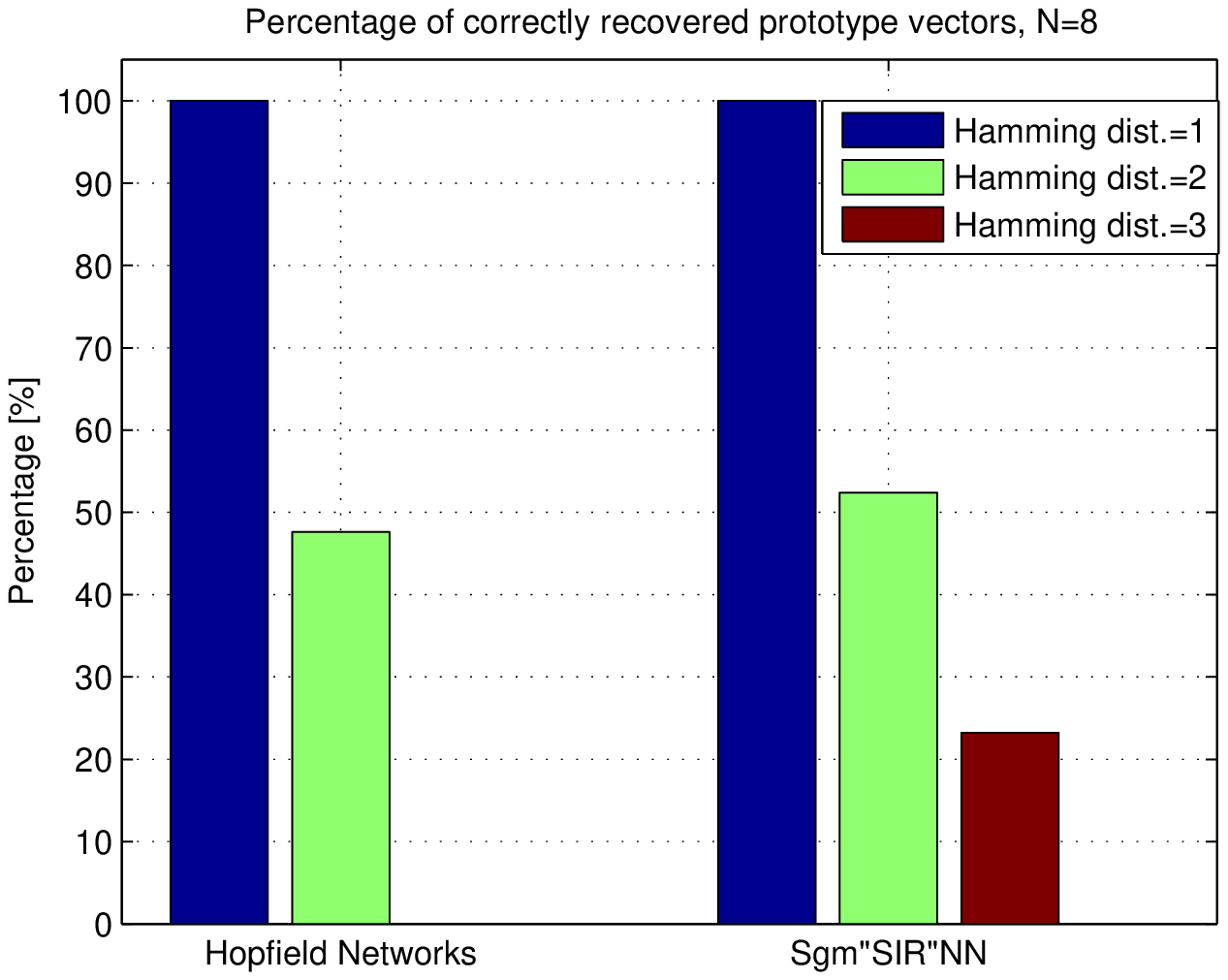}
   \vspace{-1em}        
  \end{center}
 \caption{ The figure
shows percentage of correctly recovered desired patterns for
all possible initial conditions in example 1 for the proposed Sgm''SIR''NN 
as compared to traditional Hopfield network with 8 neurons. }
\label{fig:CIRHop_ex1_percentage} 
\end{figure}

The Figure \ref{fig:CIRHop_ex1_percentage} shows the percentages of correctly recovered desired patterns for 
all possible initial conditions $\mathbf{ x }(t=0) \in (-1,+1)^N$, in the proposed Sgm"SIR"NN 
as compared to traditional Hopfield network.  

Let $m_d$ show the number of prototype vectors and $C(N,K)$, (such that $N \geq K \geq 0$), represent the 
combination $N, K$, which is equal to $C(N,K)=\frac{N!}{(N-K)! K!}$, where $!$ shows factorial. 
In our simulation, the prototype vectors are from $(-1,1)^N$ as seen above. For initial conditions,  
we alter the sign of $K$ states where $K$=0, 1, 2, 3 and 4, which means the initial condition 
is within $K$-Hamming distance from the corresponding prototype vector. 
So, the total number of different possible combinations for the initial conditions for this example is 
24, 84 and 168 for 1, 2 and 3-Hamming distance cases respectively, which 
could be calculated by $m_d \times C(8,K)$, where $m_d =3$ and $K=$ 1, 2 and 3. 



As seen from Figure \ref{fig:CIRHop_ex1_percentage}, the performance of the proposed network Sgm"SIR"NN 
is the same as that of the continuous Hopfield Network for 1-Hamming distance case ($\%100$ for both networks) and 
is slightly and noticeably higher than that of the Hopfield Network for 2 and 3-Hamming distance cases respectively. 
However, it's known that the performance of Hopfield network may highly depend on the weight matrices. 
For example, it's observed that for ${\mathbf A} = - 3{\mathbf I}$, 
the performance of Hopfield Network is slightly better than the proposed network for the same 
weights ${\mathbf W}$ and ${\mathbf A}$. 
So, our test simulation results
suggest that the proposed network Sgm''SIR''NN and the Hopfield network, in general, 
gives comparable performances in many cases. To investigate when either one ourperforms the other one 
would be an interesting future research item.

\vspace{0.2cm}
\emph{Example 2:}
\vspace{0.2cm}

The desired prototype vectors are 

\begin{equation} \label{eq:ex1_D}
{\mathbf D} =
\left[
\begin{array}{c c c c c c c c c c c c c c c c}
1 & 1 & 1 & 1 & 1 & 1 & 1 & 1 & -1 & -1 & -1 & -1 & -1 & -1 & -1 & -1 \\
1 & 1 & 1 & 1 & -1 & -1 & -1 & -1 & 1 & 1 & 1 & 1 & -1 & -1 & -1 & -1 \\
1 & 1 & -1 & -1 & 1 & 1 & -1 & -1 & 1 & 1 & -1 & -1 & 1 & 1 & -1 & -1 \\
1 & -1 & 1 & -1 & 1 & -1 & 1 & -1 & 1 & -1 & 1 & -1 & 1 & -1 & 1 & -1
\end{array}
\right]
\end{equation}

The weight matrices ${\bf A}$ and ${\bf W}$ and 
threshold vector ${\bf b}$ is obtained as follows 
by using the outer products based design as explained above. 
For matrix ${\bf A}$, $\rho$ is chosen as -2.
The other network paramaters 
are chosen as in example 1: $\sigma_1=10$, 
$\sigma_2=2$. 

\begin{eqnarray} \label{eq:matA_W_b_ex2}
{\mathbf A} & = & 2 {\mathbf I}, \nonumber \\
{\mathbf W} & = &
\left[
\begin{array}{c c c c c c c c c c c c c c c c}
 0  &  2  &  2  &  0  &  2  &  0  &  0  & -2  &  2  &  0  &  0  & -2  &  0  & -2  & -2  & -4 \\
 2  &  0  &  0  &  2  &  0  &  2  & -2  &  0  &  0  &  2  & -2  &  0  & -2  &  0  & -4  & -2 \\
 2  &  0  &  0  &  2  &  0  & -2  &  2  &  0  &  0  & -2  &  2  &  0  & -2  & -4  &  0  & -2 \\
 0  &  2  &  2  &  0  & -2  &  0  &  0  &  2  & -2  &  0  &  0  &  2  & -4  & -2  & -2  &  0 \\
 2  &  0  &  0  & -2  &  0  &  2  &  2  &  0  &  0  & -2  & -2  & -4  &  2  &  0  &  0  & -2 \\
 0  &  2  & -2  &  0  &  2  &  0  &  0  &  2  & -2  &  0  & -4  & -2  &  0  &  2  & -2  &  0 \\
 0  & -2  &  2  &  0  &  2  &  0  &  0  &  2  & -2  & -4  &  0  & -2  &  0  & -2  &  2  &  0 \\
-2  &  0  &  0  &  2  &  0  &  2  &  2  &  0  & -4  & -2  & -2  &  0  & -2  &  0  &  0  &  2 \\
 2  &  0  &  0  & -2  &  0  & -2  & -2  & -4  &  0  &  2  &  2  &  0  &  2  &  0  &  0  & -2 \\
 0  &  2  & -2  &  0  & -2  &  0  & -4  & -2  &  2  &  0  &  0  &  2  &  0  &  2  & -2  &  0 \\
 0  & -2  &  2  &  0  & -2  & -4  &  0  & -2  &  2  &  0  &  0  &  2  &  0  & -2  &  2  &  0 \\
-2  &  0  &  0  &  2  & -4  & -2  & -2  &  0  &  0  &  2  &  2  &  0  & -2  &  0  &  0  &  2 \\
 0  & -2  & -2  & -4  &  2  &  0  &  0  & -2  &  2  &  0  &  0  & -2  &  0  &  2  &  2  &  0 \\
-2  &  0  & -4  & -2  &  0  &  2  & -2  &  0  &  0  &  2  & -2  &  0  &  2  &  0  &  0  &  2 \\
-2  & -4  &  0  & -2  &  0  & -2  &  2  &  0  &  0  & -2  &  2  &  0  &  2  &  0  &  0  &  2 \\
-4  & -2  & -2  &  0  & -2  &  0  &  0  &  2  & -2  &  0  &  0  &  2  &  0  &  2  &  2  &  0 
\end{array}
\right],   \nonumber \\
{\mathbf b} & = & {\mathbf 0} 
\end{eqnarray}

\begin{figure}[htbp]
  \begin{center}
   \epsfxsize=30.0em    
\leavevmode\epsffile{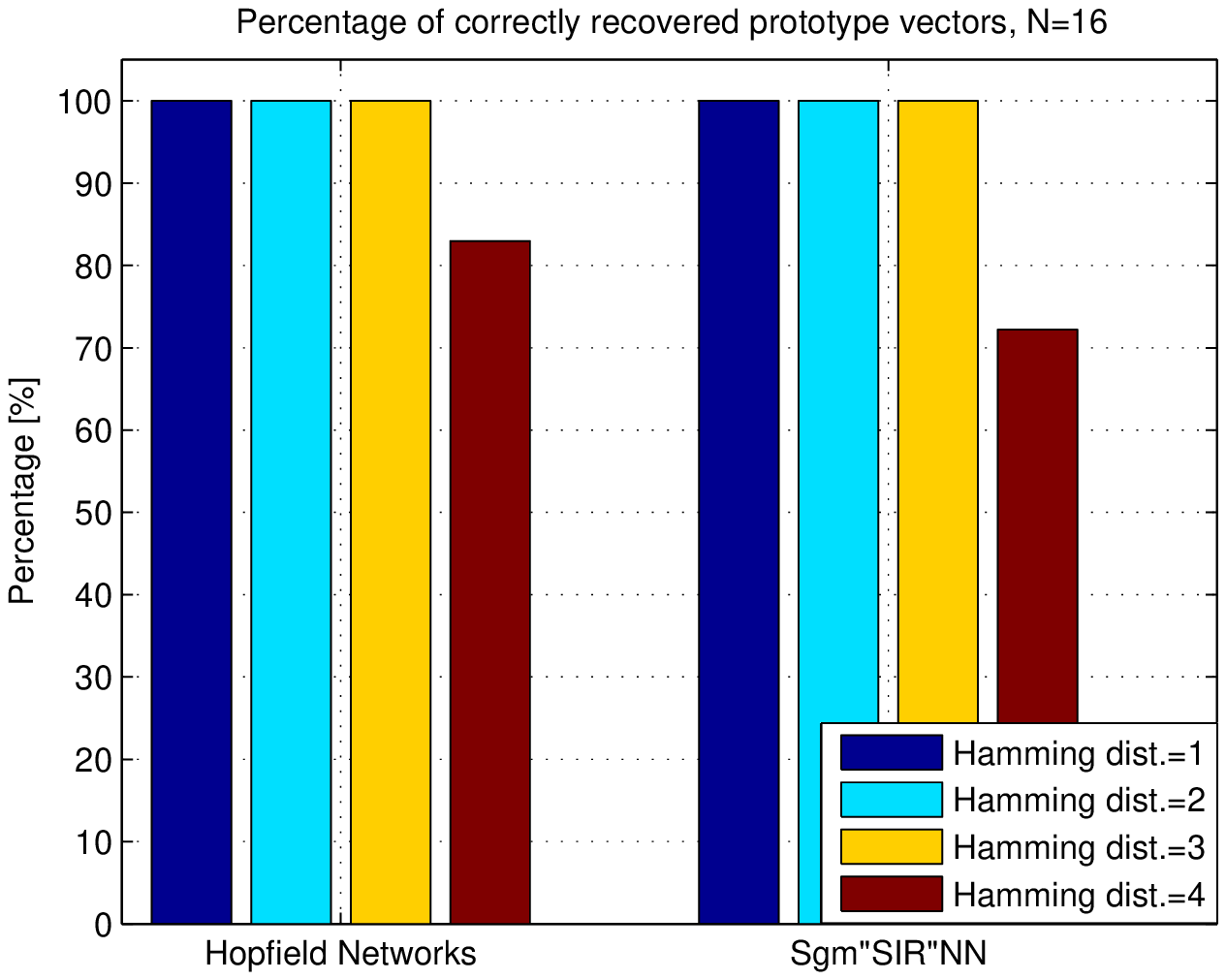}
   \vspace{-1em}        
  \end{center}
 \caption{ The figure 
shows percentage of correctly recovered desired patterns for 
all possible initial conditions in example 2 for the proposed Sgm"SIR"NN
as compared to traditional Hopfield network with 16 neurons. }
\label{fig:CIRHop_ex2_percentage} 
\end{figure}

The Figure \ref{fig:CIRHop_ex2_percentage} shows the percentages of correctly recovered desired patterns for 
all possible initial conditions $\mathbf{ x }(t=0) \in (-1,+1)^{16}$, in the proposed Sgm"SIR"NN 
as compared to traditional Hopfield network.  

The total number of different possible combinations for the initial conditions for this example is
64, 480 and 2240 and 7280 for 1, 2, 3 and 4-Hamming distance cases respectively, which
could be calculated by $m_d \times C(16,K)$, where $m_d =4$ and $K=$ 1, 2, 3 and 4. 

As seen from Figure \ref{fig:CIRHop_ex2_percentage} the performance of the proposed network D-Sgm"SIR"NN
is the same as that of Hopfield Network for 1, 2 and 3-Hamming distance cases ($\%100$ for both networks).  
The Hopfield network network gave slightly better performance than the proposed network 
for 4-Hamming distance case.

\section{Concluding Remarks  \label{Section:CONCLUSIONS}}

In this paper, starting from the power control algorithm in \cite{Uykan04}, 
we present a Sigmoid-based ``Signal-to-Interference Ratio, (SIR)'' balancing
dynamic network, called Sgm''SIR''NN, which includes both the Sigmoid power 
control algorithm (SgmDPCA) and the 
Hopfield neural networks, two different areas whose scope of interest, motivations and settings 
are completely different. The stability of the Sgm"SIR"NN is examined by the proposed Lyapunov functions.

Starting from the differential equation form of the Sigmoid DPCA in 
\cite{UykanPhD01} and \cite{Uykan04} and relaxing the
strick restrictions and assumptions on the positiveness and spectral radius of the link gain matrix,
we establish a link from SgmDPCA to the Hopfield-like NNs. The proposed approach yields a Sigmoid basis
SIR-balancing NN which exhibits similar features as Hopfield NN does. 
Computer simulations show the effectiveness of the
proposed network as compared to traditional Hopfield Network.

Our investigations show that
1) The proposed Sgm''SIR''NN includes both SgmDPCA algorithm and Hopfield NN as special cases.
2) The Sgm''SIR''NN exhibits features which are generally attributed to Hofield-like recurrent NN.
3) Estalishing an analogy to the SgmDPCA, the proposed network as well as the Hopfield Network
keeps the fictitious SIR at a target level.

As a continuation of this work, 
we examine the proposed network in discrete time and compare it to 
discrete Hopfield Networks in \cite{Uykan08b}.

\vspace{0.2cm}

\section*{Appendix}

In what follows, we will show the sigmoid function 
($f(a) = 1 - \frac{2}{1 + exp(-\sigma a)}, \quad \sigma>0$) has the global Lipschitz constant
$k = 0.5 \sigma$.

Since $f ( \cdot )$ is a differentiable function, we can apply the mean value theorem 

\begin{eqnarray} \label{eq:nebiliim}
 f(a) - f(b) = (a-b) f^{'}( \mu a + (1- \mu) (b-a))  \nonumber \\
with \quad \mu \in [0,1]  \nonumber
\end{eqnarray}

The derivative of $f(\cdot)$ is $f^{'}(a) = \frac{\sigma}{e^{\sigma a} {(1+e^{\sigma a})^2} }$ 
whose maximum is at the point $a = 0$, i.e., $| f^{'}(a) | \leq 0.5 \sigma$.  So we obtain the 
following inequality 

\begin{equation}\label{eq:atkafadan}
| f(a) - f(b) | \leq k |a-b|  
\end{equation}

where $k = 0.5 \sigma$ is the global Lipschitz constant of the sigmoid function.  

\vspace{0.2cm}


\vspace{0.2cm}

\section*{Acknowledgments}

This work was supported in part by Academy of Finland and Research Foundation (Tukis\"{a}\"{a}ti\"{o}) of Helsinki 
University of Technology, Finland.

\nocite{*}
\bibliographystyle{IEEE}

%

\vspace{3cm}



\newpage

\listoffigures

\newpage

\newpage

\newpage

\end{document}